\documentclass[final]{aipproc}

\layoutstyle{6x9}

\def\apj{Astrophysical Journal}

\def\aap{Astronomy \& Astrophysics}

\def\araa{Annual Rev. of Astronomy. \& Astrophysics}
\def\mnras{Monthly Notices of the Royal Astronomical Society}

\def\apjl{Astrophysical Journal Letters}
\def\pasp{Publications of the Astronomical Society of the Pacific}
\begin{document}

\title{Accretion flows around stellar mass black holes and neutron stars}

\author{Didier Barret\footnote{send email to Didier.Barret@cesr.fr}}{
  address={Centre d'Etude Spatiale des Rayonnements, CNRS/UPS, 9 Avenue du
Colonel Roche, 31028 Toulouse Cedex 04, France} }

\begin{abstract} In this review, I summarize the main X-ray/hard X-ray properties of
the accretion flows around black holes and neutron stars based on recent broad-band
spectral and timing observations performed by the BeppoSAX and Rossi X-ray Timing
Explorer satellites. Emphasizing the spectral and timing similarities observed between
black holes and neutron stars, I discuss on the most likely accretion geometry and emission processes associated with hard and soft spectral states. For black holes, in the hard state, the accretion geometry is more likely made of a truncated disk and a hot inner flow, in which thermal Comptonization takes place. The truncated disk is likely to be the dominant source of seed photons, and the site for the production of the reflection component observed. In soft states, the disk now extends closer to the compact object and is brighter in X-rays. The hard X-ray emission occurs through Comptonization of disk photons on a thermal/non-thermal electron distribution, generated in magnetic flares above the accretion disk. For neutron stars, similar accretion geometry and emission mechanisms may apply but the unavoidable radiation from the neutron star surface adds yet another component in the X-ray spectrum. It also acts as an additional source of cooling for the Comptonizing cloud, leading to softer spectra in neutron stars than in black holes.
\end{abstract}

\maketitle

\section{Introduction} Accretion is a very common process in the Universe
playing a key role from young stellar objects to super-massive black holes
powering active galactic nuclei. Accretion of matter onto a neutron star or a
stellar mass black hole will produce copious X-ray emission (\cite{fkr02} and references therein). In most X-ray
binaries, the matter from the secondary star feeds the compact object
through an accretion disk. In the disk, matter heats up as it approaches the central object in keplerian orbits, while viscosity transports angular momentum outward. The energy spectrum and time variability of the X-rays can be used to probe the strong gravity field of the collapsed star, the accretion geometry and the emission processes at work under extreme conditions.

Recent advances in understanding accretion flows around compact objects have
greatly benefited from the observations performed by two exceptional X-ray
observatories: BeppoSAX \cite{frontera98} and RXTE \cite{bradt93aa}. The major
BeppoSAX contribution came from its broad band capability (from less than a keV to $\sim
200$ keV) and unprecedented sensitivity in the hard X-ray range (E$ \ge 40$
keV), whereas the RXTE contribution came from its ability to perform extensive and flexible monitoring spectral and timing observations.

This review presents a summary of what has been learned recently from these observations by combining spectral and timing information and emphasizing the many similarities observed between neutron stars and black holes. The X-ray properties of accreting binaries have been reviewed extensively in the last few years \cite{donereview01,belo01review,poutanen01review,natalucci01,donereview02,mcre03cup,belloni03,disalvostella03}. The optical, ultraviolet and infrared observations of X-ray binaries
are reviewed in \cite{charles03}, whereas their radio properties are reviewed in \cite{fender03}. 
%
%This review is slightly biased towards neutron star binaries, because significant progresses have been recently accomplished in our understanding of accretion flows around compact objects through the studies of a few of these systems \cite{GDBSAXJ-02,GD1608-03,OBG1705-03,BO1705-03,VVM1608-03}.

Since this review is also intended for plasma physicists not always familiar with the terminology used in the field, it is worth defining precisely what we are talking about. With neutron star binaries, it is meant those systems containing weakly magnetized neutron stars (B$<10^9$ G); the presence of a neutron star is inferred from the detection of type I X-ray bursts (thermonuclear flashes on the neutron star surface) or X-ray pulsations (for accreting millisecond pulsars). There are more than 50 such systems known in the Galaxy \cite{liu01}. Similarly black hole binaries mean those systems for which there are compelling evidence that the compact object is a black hole (e.g. estimate of the mass function which indicates a compact object mass larger than $\sim 3$ solar masses: the maximum mass of a stable neutron star \cite{mcre03cup}). To date, there are 18 such systems \cite{mcre03cup}, among the $\sim 150$ X-ray binaries known. 

%There is a great variety of behaviors observed within and between objects. In order to keep the picture as clear as possible, I will therefore limit myself to the most common and well established observational facts. I will not enter into the details of the zoology of the spectral and timing states of accreting X-ray binaries, which sometimes complicates the picture in the extreme. Still, it is impossible to review the X-ray properties of accreting binaries without defining the two main spectral states in which most of them are observed: the soft and hard spectral states.

\begin{figure}[!t]
\includegraphics[width=2.8in]{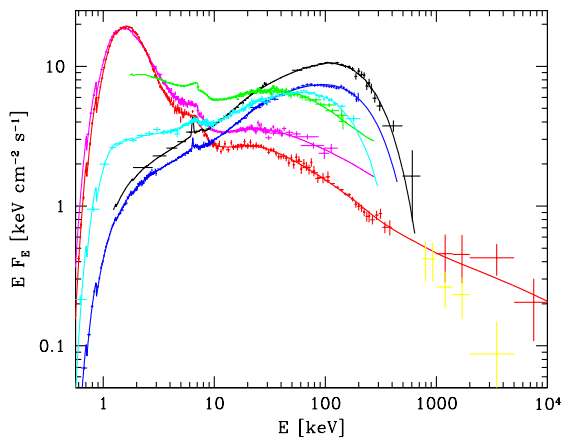}
\caption{Example broadband spectra of the black hole Cyg X-1 from pointed observations. (Courtesy of Andrzej Zdziarski from \cite{zdz02}). In the soft spectral state, the peak of the energy is radiated below 10 keV, and a non-thermal component extends up to 1 MeV. In the hard spectral state, the peak of the energy is radiated above 10 keV. The energy cutoff at $\sim 150$ keV is the signature of thermal Comptonization.}
\label{barret_fig1}
\end{figure}

\section{Soft and hard spectral states} 
X-ray binaries for which the accretion rate onto the compact object varies exhibit various spectral states. This is naturally the case for X-ray transients, which sample a very wide range of accretion rates from quiescence to outburst. For the sake of clarity, in this paper I will define soft and hard spectral states based on the fraction of luminosity radiated below 10 keV\footnote{10 keV has been chosen to encompass the peak of the energy spectra of the persistently bright X-ray sources which are permanently observed in soft spectral states, e.g. Sco X-1.}. In a soft spectral state, the bulk of the energy (e.g.
$\sim 80$\%) is radiated below 10 keV, whereas in a hard spectral state, the
energy radiated below 10 keV is either comparable or less than the energy radiated above. There are obviously intermediate and extreme states beside soft and hard states (see Figure \ref{barret_fig1}). In general, it is assumed that the X-ray luminosity (below 10 keV) is a good measure of the source bolometric luminosity, and hence that the source is more luminous in a soft state than in a hard state; but this is not necessarily the case. Thanks to RXTE, many systems have now been followed extensively during their spectral state transitions, and the picture that is emerging is that the spectral state is a very complex function of the source luminosity (e.g. \cite{belloni03, mcre03cup,barret01review}). The luminosity associated with spectral state transitions vary within and between systems \cite{mcre03cup}. Within a system, it depends also to some degree on the long term evolution of the accretion rate \cite{BO1705-03}.

In the following, we review separately each of the spectral components making up the broad band X-ray/hard X-ray spectrum of accreting X-ray binaries.
\section{Thermal and non-thermal Comptonization}
\begin{figure}[!t]
\includegraphics[width=3.0in]{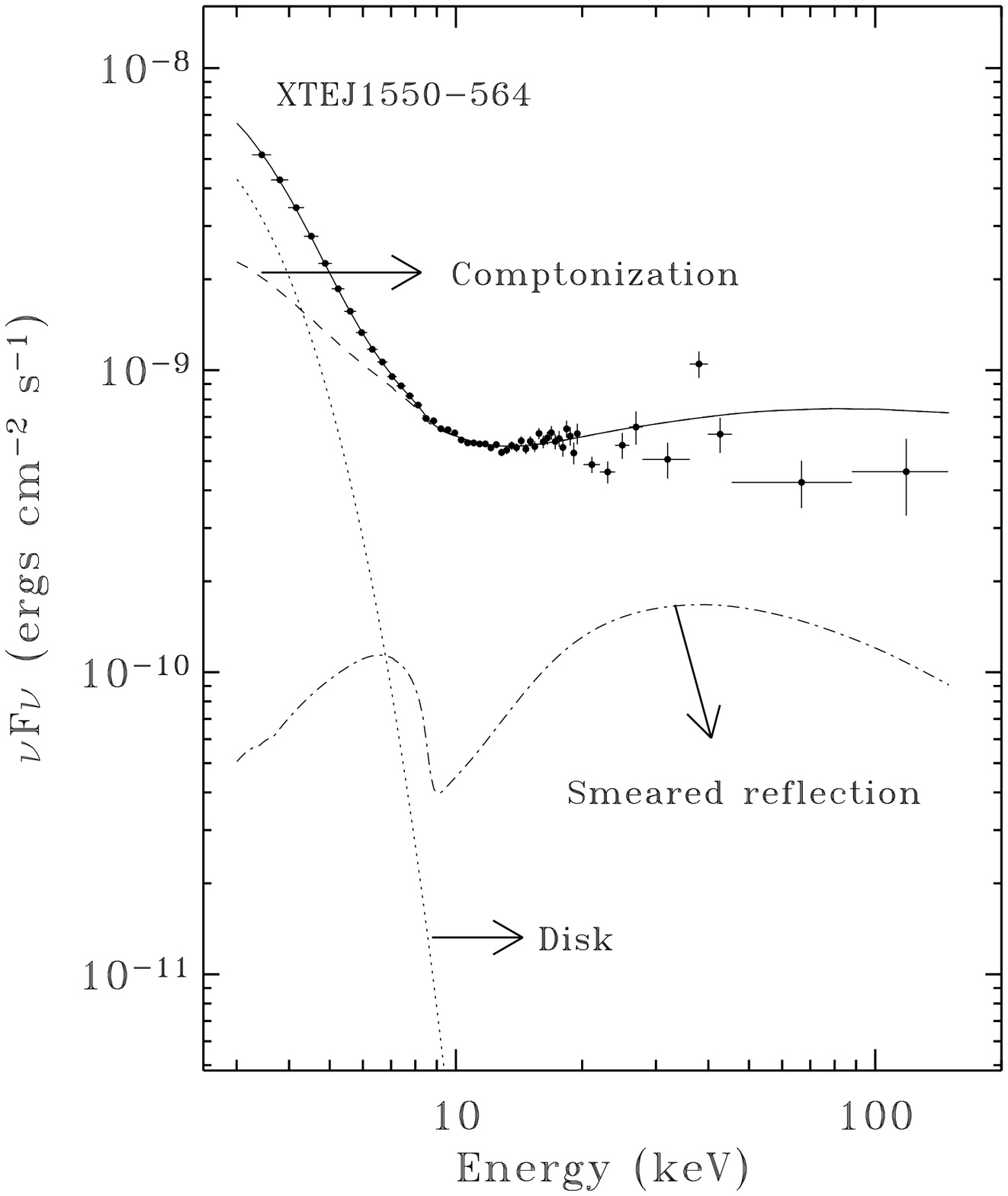}\hspace*{-1.8cm}\includegraphics[width=3.0in]{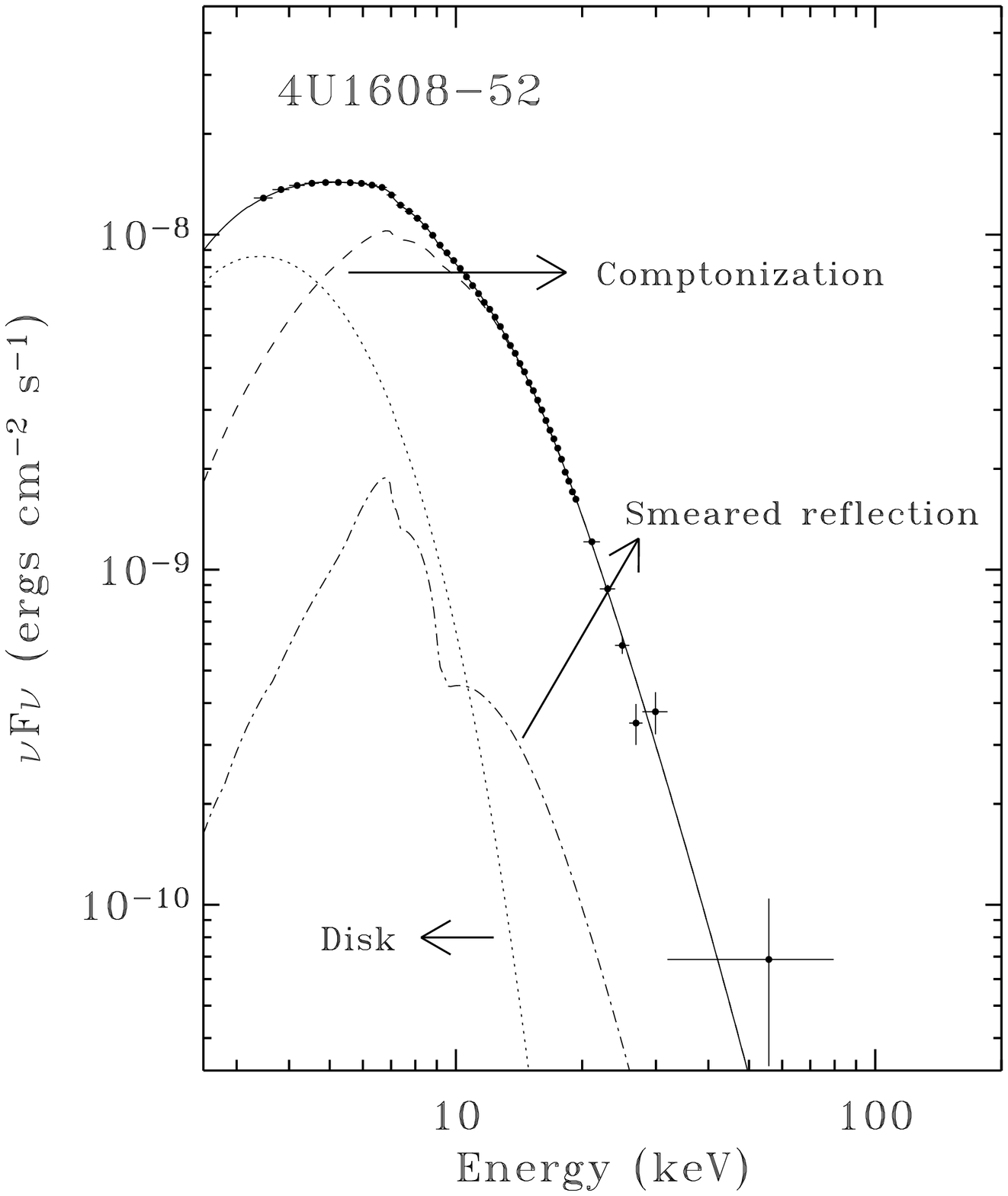}
\caption{{\it left)} Soft state spectrum of the black hole system XTEJ1550-564. {\it right)} Soft state spectrum of the neutron star system 4U1608-52 (both data points are courtesy of Marek Gierli{\' n}ski, and are taken from \cite{DGhorizon-03,GD1608-03,GD1550-03}).}
\label{barret_fig2}
\end{figure}

\begin{figure}[!t]
\includegraphics[width=3.0in]{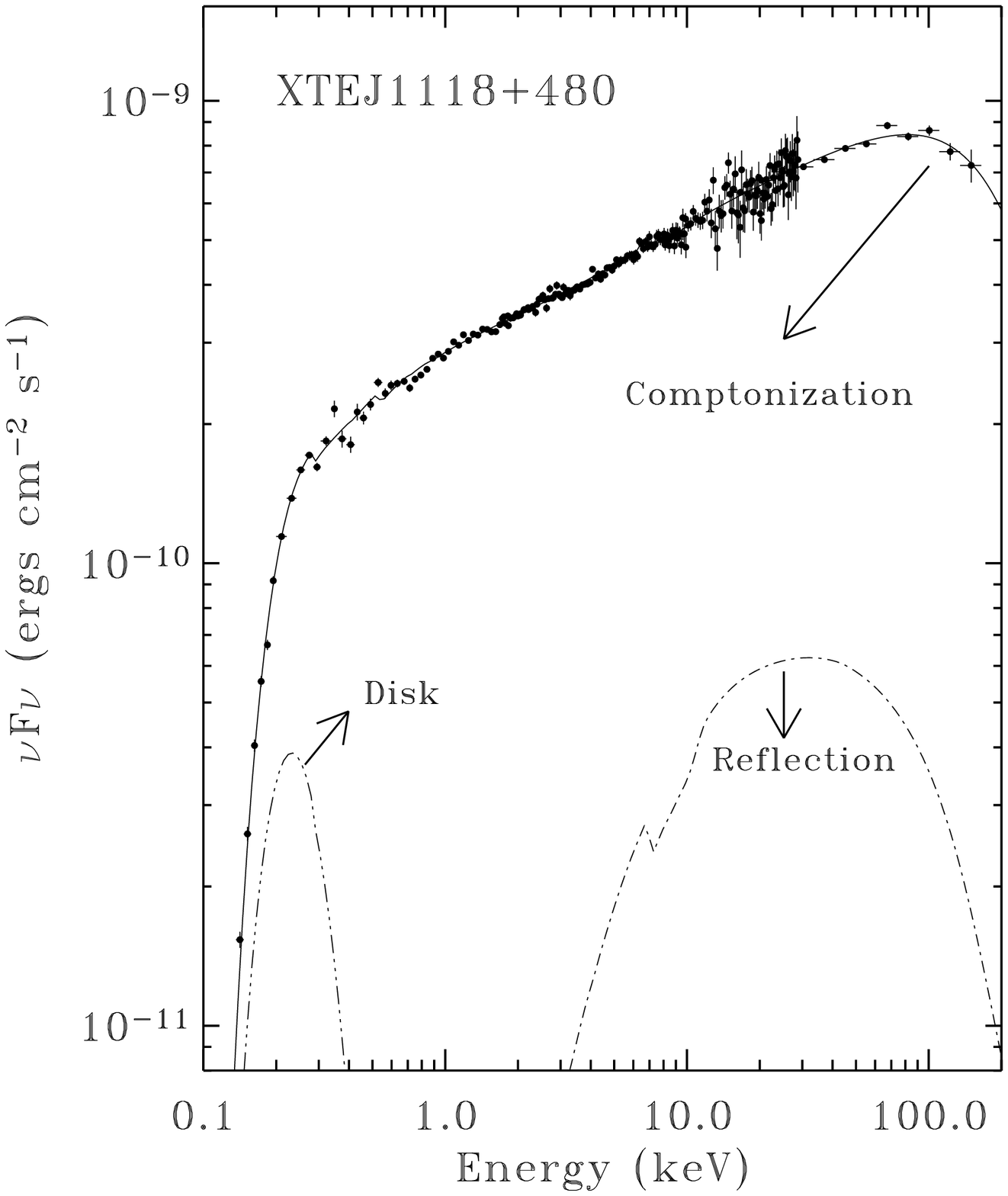}\hspace*{-1.8cm}\includegraphics[width=3.0in]{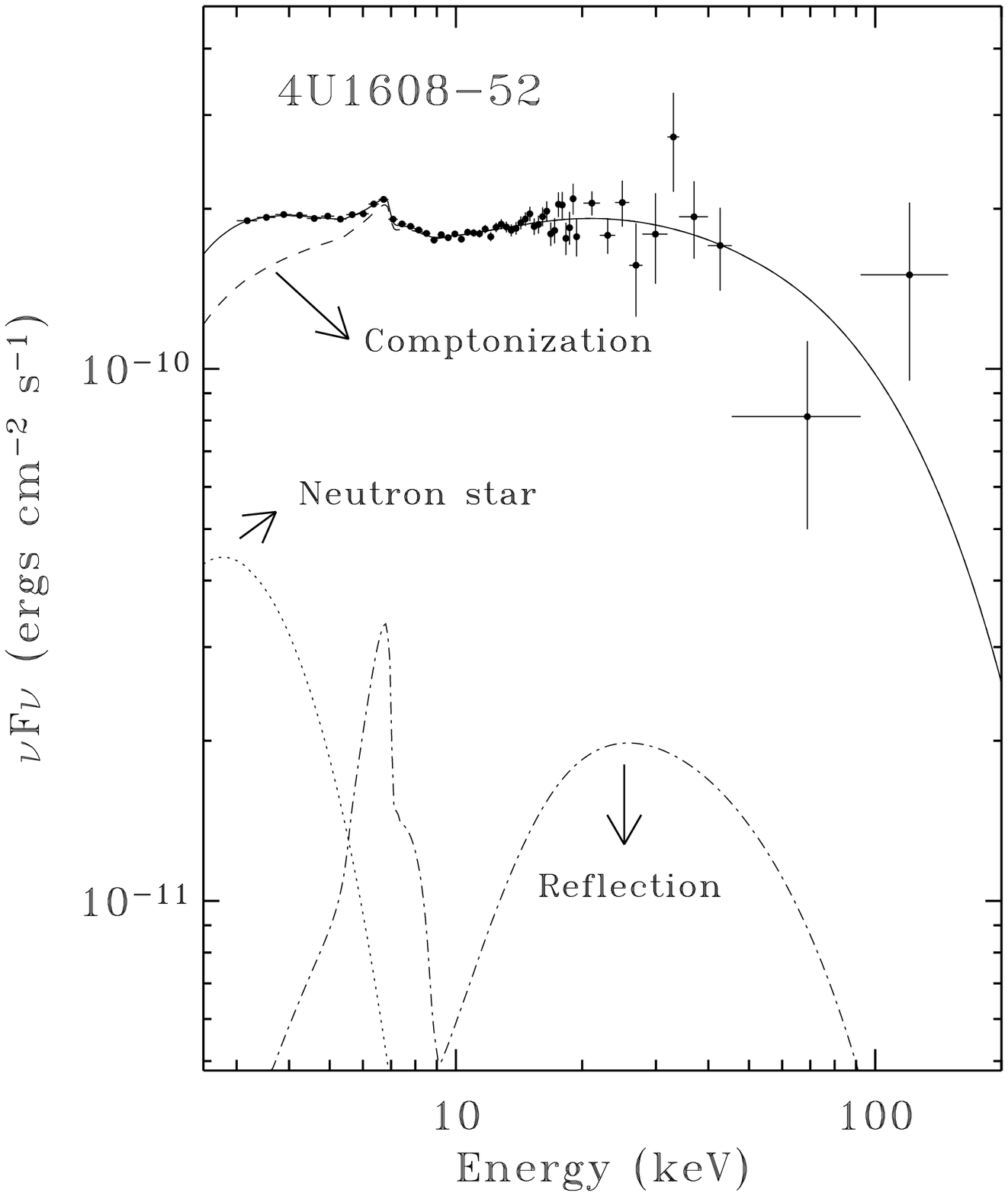}
\caption{{\it left)} Hard state BeppoSAX spectrum of the black hole XTEJ1118+480 (Courtesy of Filippo Frontera and Lorenzo Amati from \cite{frontera1118-03}). {\it right)} Hard state RXTE spectrum of the neutron star 4U1608-52 (Courtesy of Marek Gierli{\' n}ski from \cite{GD1608-03}).}
\label{barret_fig3}
\end{figure}

Comptonization, which is the process of Compton up scattering of soft seed photons on hot electrons is the dominant emission mechanism in both soft and hard spectral states of X-ray binaries\footnote{This is not the case however for the ultrasoft state of black holes when the disk emission dominates over a weak comptonized hard tail.} (see Figures \ref{barret_fig2} \& \ref{barret_fig3}). The shape of the Comptonized spectrum depends primarily on the nature of the electron distribution (thermal/non-thermal), the temperature and optical depth of the electron cloud, the characteristic temperature of the seed photons, and to a lower degree on the geometrical shape of the scattering cloud (e.g. \cite{titar94}).

Thermal Comptonization is observed in black hole binaries in their hard states,
and both in soft and hard states of neutron star binaries. For black holes,
their broad band spectra show clear evidence for a high energy cutoff (the signature of a thermal electron distribution) around 200 keV (see Figures \ref{barret_fig1} \& \ref{barret_fig3}). Fitting these spectra with thermal Comptonization models yields for the electron plasma a temperature ranging from $50$ to $\sim100$ keV and an optical depth of the order of unity (e.g. Cyg X-1, see \cite{disalvocyg-01}, but see also \cite{frontera01} for a more sophisticated model involving two temperatures for the electrons). 

For neutron stars, in soft states, the electron temperature is typically 3 to 5 keV and the optical depth of the electron plasma a few \cite{barret01review}. On the other hand, in their low states, the comptonized spectra are considerably harder and hotter, with the electron temperature typically around $\sim 30$ keV \cite{barret01review,disalvostella03} (see Figures \ref{barret_fig2} \& \ref{barret_fig3}). These spectra share clearly some similarities with those observed from hard state black hole binaries. However on average, neutron stars reach electron temperature lower than black holes. This has been interpreted as the signature of the neutron star surface acting as a Compton thermostat to limit the electron cloud temperature (see discussion in \cite{barret01review}). 

Non-thermal Comptonization has been observed in soft states of black holes (Figure \ref{barret_fig1}). Significant emission is detected up to $\sim 1$ MeV and the spectra show no clear evidence for a high energy cutoff \cite{grove98}. These spectra have been
adequately fitted with Comptonization models assuming a hybrid thermal and
non-thermal electron distribution. A significant fraction (up to $\sim50\%$) of
the power must go in the non-thermal electrons to account for the spectrum
beyond $\sim 300$ keV. The most likely mechanism to power the non-thermal electrons is magnetic reconnection in a corona above the accretion disk (e.g. \cite{GD1550-03}).  

For neutron star binaries, the evidence for non-thermal Comptonization comes from the observation of non-attenuated power law tails. Such tails have been observed from some of the brightest neutron star binaries, permanently observed in soft states (e.g. \cite{disalvogx-01}). These tails dominate the spectrum above 30 keV and carry only a small fraction of the source luminosity (a few \%). Non-attenuated power laws with slopes in the range 2.0-2.5 have also been observed from much fainter systems \cite{natalucci01,disalvostella03}. Fitting such spectra with Comptonization models sets a lower limit on the electron temperature to at least 100 keV. 

The origin of these tails is currently debated but may be related to the non-thermal components observed in the soft spectral states of black hole binaries \cite{disalvostella03}. In bright neutron stars, the luminosity in the tail remains a small fraction of the source luminosity. This is not the case for black holes. This could again be related to the additional Compton cooling from the neutron star surface/boundary layer emission \cite{DGhorizon-03}. Alternatively, it has also been suggested that the tails may be related to the appearance of a jet (which is observed in some bright systems, e.g. \cite{fender03}). The high energy tail would then result either from non-thermal Comptonization of disk seed photons in the jet or from direct optically thin synchrotron emission of the jet itself (see discussion in \cite{disalvostella03}).

\section{Seed photons} 
In the section above, I have left unspecified the source of seed photons for the
Comptonization. These seed photons are sometimes directly visible in the X-ray spectrum in the form of a soft component. This component is generally modeled by a multi-color disk blackbody model (e.g. \cite{merloni00}), and provides some constraints on the geometry of the accretion flow.
 
In black holes binaries, especially in their soft states, it has been shown that the soft component was consistent with being the unscattered fraction of the seed photons supplied by the accretion disk to a magnetic corona \cite{GD1550-03} (Figure \ref{barret_fig2} left). The changes in the ratio between the disk and corona emissions are explained by changing the optical depth of the comptonizing cloud and/or the covering factor of the disk by the corona. In the highest luminosity states, the disk emission is so strongly comptonized that it becomes barely visible in the X-ray spectrum (e.g. \cite{donereview01}).

When going to hard states, the soft component generally becomes cooler (hence fainter in X-rays). One of the best example to date is provided by the black hole binary XTEJ1118+480, whose extremely low interstellar absorption allowed to measure the soft component temperature down to $\sim 40$ eV \cite{frontera1118-03}; its hard state spectrum is shown in Figure \ref{barret_fig3} (left). If attributed to the accretion disk, the soft component would imply that the disk is truncated at $R\ge30$ R$_g$ (with $R_g=GM/c^2$). Near-simultaneous infrared to hard X-ray observations of the same source also in a hard state resolved the disk emission into an even softer 24 eV thermal component, from which an inner disk radius larger than 70 $R_g$ was inferred \cite{mcclintock01}. 

Three classes of models are currently competing to explain hard state spectra  (see \cite{donereview01,beloreview01} for a review). The first one, is the so-called disk-spheroid model (e.g. \cite{poutanen-97}), in which a hot inner flow is surrounded, with a partial overlap, by a cool accretion disk supplying the seed photons. The hot inner flow is a two-temperature plasma, with the protons much hotter than the electrons. The two-temperature plasma cools either by Compton scattering (e.g. \cite{shapiro76}) or by advection (e.g. \cite{esin-01}). In the second class of model, electrons are stored in magnetic flares above a non truncated accretion disk. These flares could be generated by the magneto-hydrodynamic dynamo responsible for the disk viscosity \cite{hawley00}. The third one uses the fact that hard states of black holes are associated with compact and quasi-steady radio jets \cite{fender03}. It has been suggested that the X-rays could arise directly from synchrotron emission in the jet (e.g. \cite{markoff03}).

Sometimes the soft component detected in the spectrum is so weak that it is no longer consistent with being the visible part of the seed photon spectrum \cite{frontera1118-03}. Thus alternative interpretations for the source of seed photons have been considered; as for instance thermal synchrotron emission from within the electron plasma cloud \cite{wzsyn-00}.

For neutron star binaries, in addition to the disk, thermalized emission at the neutron star surface can also supply seed photons. Since both components can be visible in the X-ray spectrum (their unscattered fraction), the interpretation of the spectra is not as straightforward as in the case of black holes. However, recent RXTE monitoring observations have shown that the complex spectral behavior of these systems could be explained also in the framework of the disk-spheroid model, to which the emission from the neutron star surface must be added. In the hard spectral state\footnote{The upper horizontal branch of the so-called "island" state.}, the disk does not contribute to the X-ray spectrum, and the seed photons are predominantly provided by the neutron star surface (limited to a hot spot in the case of accreting millisecond pulsars in which the inner flow is channeled by the neutron star magnetic field, e.g. SAXJ1808-3659, \cite{GDBSAXJ-02,poutanen03}). A fraction of the neutron star (or hot spot) emission is visible through the scattering cloud which has a moderate optical depth \cite{GDBSAXJ-02,GD1608-03,BO1705-03} (see Figure \ref{barret_fig3}, right). In that state, the lack of correlation between the spectral and timing parameters provides support for this interpretation (see \cite{GDBSAXJ-02,OBG1705-03,VVM1608-03} and discussion below). 

On the other hand, in the soft spectral state, some seed photons are still provided by the neutron star surface but the contribution from the accretion disk is no longer negligible (see Figure \ref{barret_fig2} right). Only the unscattered disk emission can be seen, as the one from the neutron star surface is buried under the Comptonizing cloud of increased optical depth. The changing optical thickness of the boundary layer, the relative strength of its unscattered emission and the one from the disk would explain the complex spectral evolution of neutron stars, as followed in their X-ray color-color diagrams \cite{GD1608-03,BO1705-03}. In addition, the neutron star/boundary layer emission which is absent in black holes would explain why in some regions of these color-color diagrams (the very soft regions) only black holes can be found \cite{DGhorizon-03}.

In both soft spectral states of neutron stars, the Comptonized component carries out most of the energy (see Figures \ref{barret_fig2} and \ref{barret_fig3} right) and dominates over the disk contribution. The site of Comptonization must therefore involve the boundary layer, in which the kinetic energy of the accreted material must eventually be released. In general relativity, it is known that the luminosity of the boundary layer can exceed that of the disk by about a factor of 2 \cite{ss86}. In the hard state, the observations suggest that the boundary layer is optically thin and merge smoothly with the hot inner flow (e.g. \cite{barret-00,barret01review}).

\section{Reflection}
Reflection of hard X-rays on some material (neutral or ionized) will produce a broad X-ray bump (peaking at $\sim$ 20 keV) together with fluorescence lines and edges of the most abundant elements (mostly iron). This reflection component holds also some potential to constrain the geometry of the accretion flow in the immediate vicinity of the compact star (see \cite{fabian00,donereview01} and references therein and e.g. \cite{miller01} for an application to the candidate black hole XTE J1748-288 X-ray data). The parameters of the reflection component and associated features can be constrained from spectral fitting, but requires a precise knowledge of the continuum spectrum (hence broad band measurements).

\begin{figure}[!t]
\includegraphics[width=2.2in]{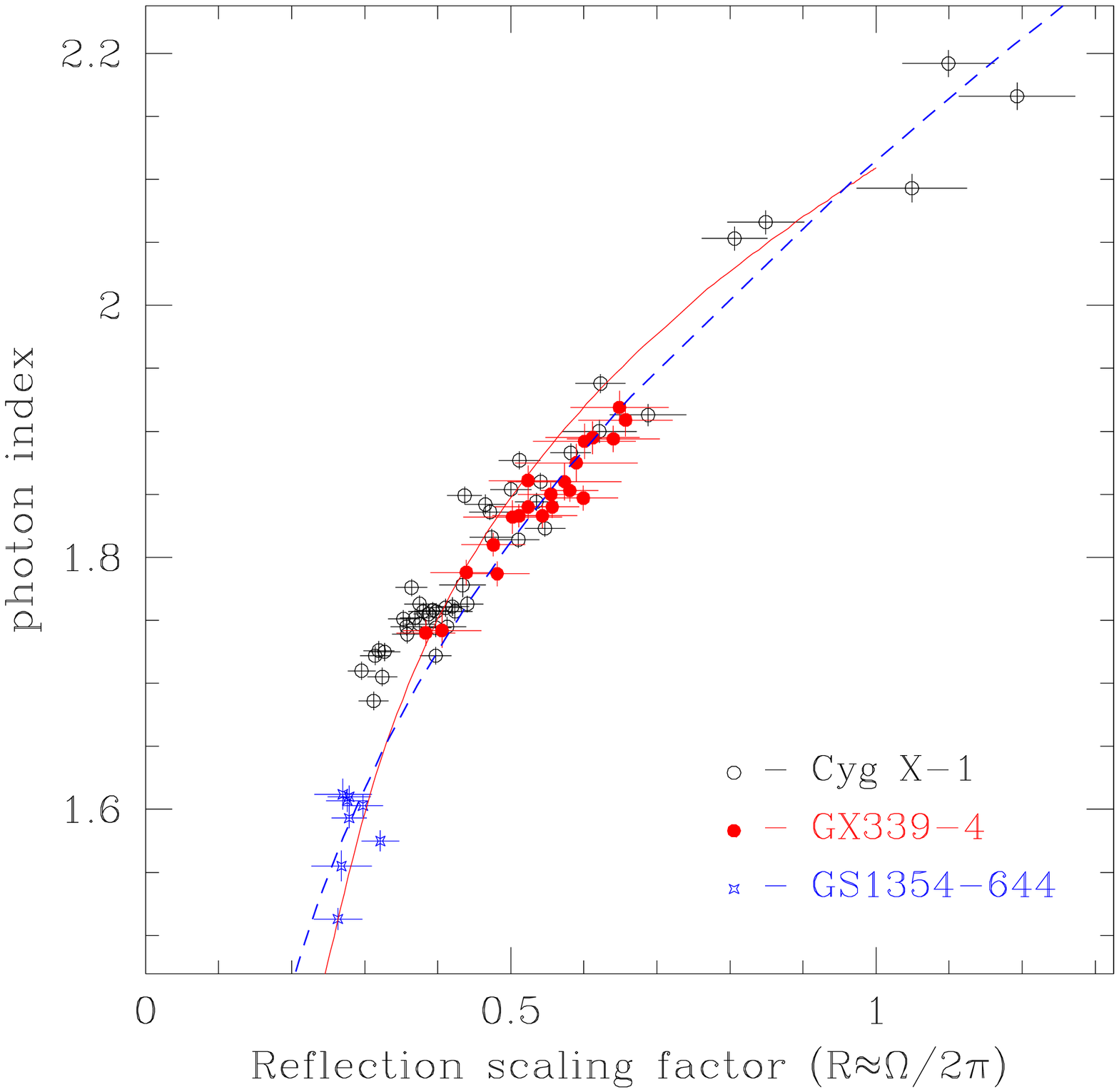}
\includegraphics[width=2.2in]{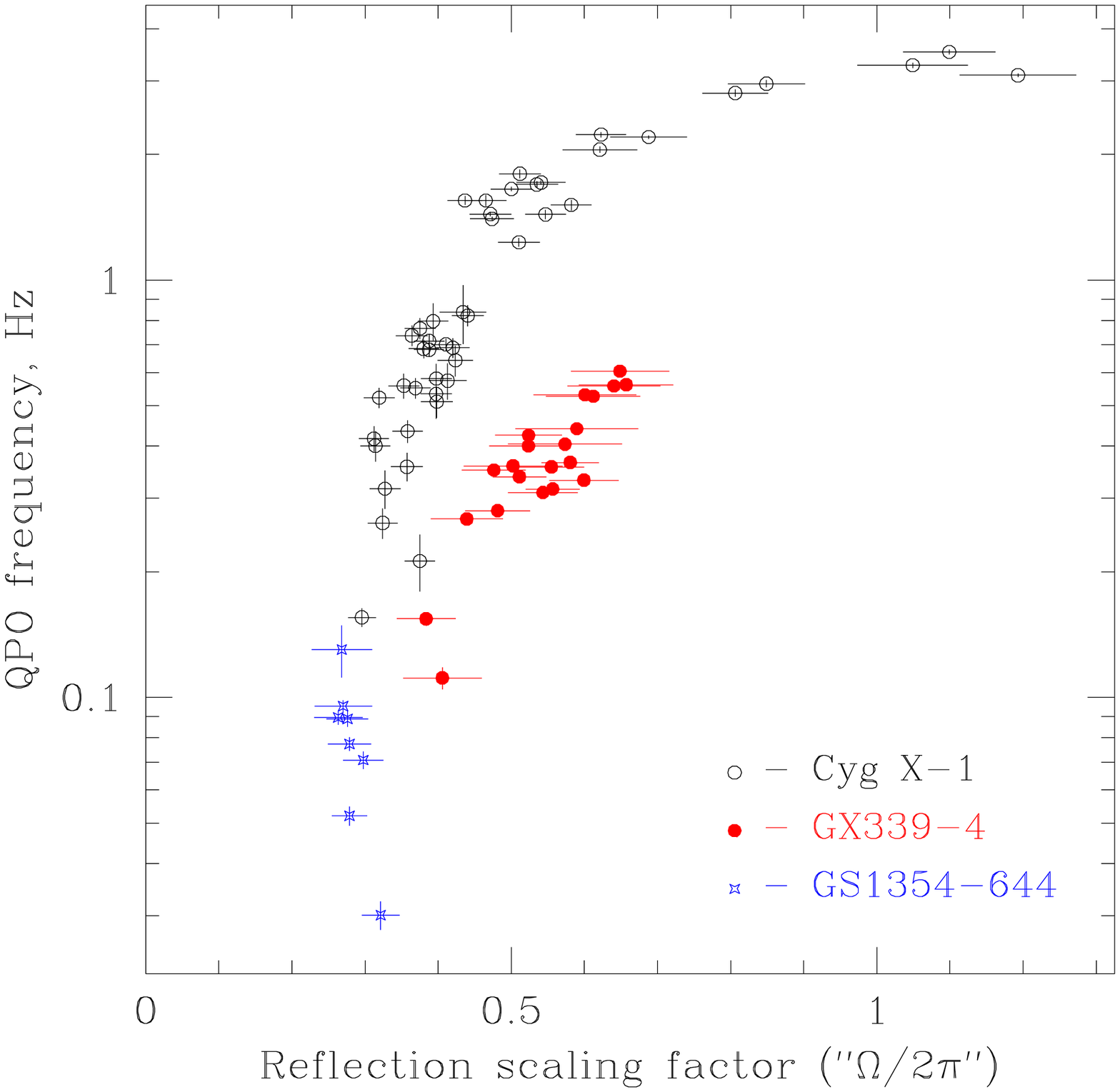}
\caption{{\it left)} Best fit value of the photon index of the underlying power law versus the reflection scaling factor for three black hole binaries. {\it right)} The QPO centroid frequency versus the relfection scaling factor for the same three objects with the same observations. (Courtesy of Marat Gilfanov from \cite{gilfanovref-00}).}
\label{barret_fig4}
\end{figure}

This reflected component has now been observed in the spectra of both neutron star
and black hole systems, in soft and hard spectral states (see Figures \ref{barret_fig2} \& \ref{barret_fig3}). However it is much easier to dig out in hard state spectra. In general, in the hard state, it is found that the reflector is mostly neutral, subtends a small solid angle much less than $2\pi$, and that the amount of relativistic smearing is much less than what would be expected if the reflector extended down to the last stable orbit \cite{donereview01}. On the other hand, in soft states, the properties of the reflection component (increased smearing) suggest that the reflector extends closer to the central object (e.g. \cite{miller01}).

The strength of the reflection component seems to be correlated with the spectral index of the continuum spectrum in a way that softer spectra are associated with more reflection \cite{zdziarski-99,gilfanovref-99,gilfanovref-00}. Fitting of the 3-20 keV spectra with a power law continuum model and its reflection yields the correlation represented in Figure \ref{barret_fig4} (left) for three black hole systems. This correlation has been interpreted as strong evidence that the reflecting material is also the source of source of seed photons, hence most likely the cool accretion disk \cite{zdziarski-99,gilfanovref-99,gilfanovref-00}. The above correlation can be interpreted in the framework of the disk-spheroid model as resulting from a geometrical effect (when the disk moves in, the solid angle it subtends to the hard source increases, leading to more reflection and more cooling, hence softer spectra) \cite{zdziarski-99,gilfanovref-99,gilfanovref-00}. Another possible interpretation of the above correlation involves emitting blobs of plasma, moving at semi-relativistic speeds above a non truncated disk \cite{beloborodov-99}. The correlation shown in Figure \ref{barret_fig4} (left) for black holes is observed also from neutron star systems, but slightly offset towards softer spectra for the same amount of reflection (Gilfanov, private communication). This could be due again to the additional Compton cooling from the neutron star surface \cite{gilfanov03}.

\section{Timing}
The Rossi X-ray Timing Explorer has opened a new window to
explore the time variability of the X-ray emission down to the dynamical time
scales of the accretion flow \cite{vanderklis00review,psaltis01,wijnands01review}. Despite great observational progresses, the origin of the variability remains poorly understood at the moment. In particular, there are no undisputed models allowing to relate the observed variability time scales to
some physical and measurable properties of the accretion flow. 

\begin{figure}[!t]
\includegraphics[width=3.0in]{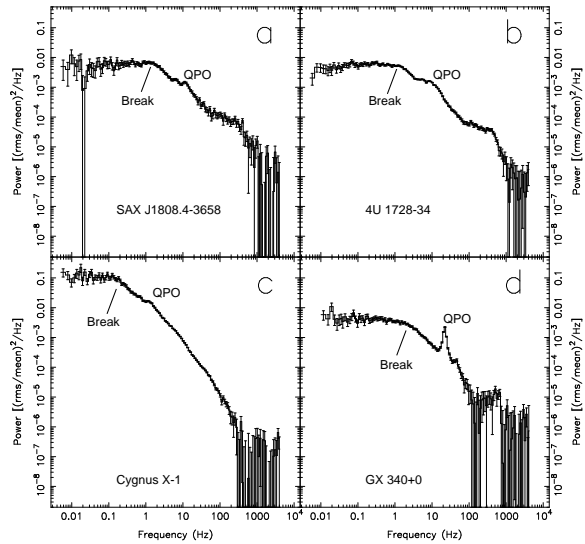}
\caption{Typical broad band power density spectra of three neutron star systems; the millisecond pulsar SAXJ1808.4-3658, 4U1728-34, GX340+0, and the black hole binary Cyg X-1. (Courtesy of Rudy Wijnands from \cite{wijnandsvanderklis-99}).}
\label{barret_fig5}
\end{figure}

Nevertheless, detailed modeling of the Fourier power density spectrum has allowed to determine and follow precisely the characteristic time scales of the variability for a variety of systems and over a wide range of source intensity states. This analysis has demonstrated, that black holes and neutron stars have very similar properties of their time variability \cite{wijnandsvanderklis-99}. The similarities are very pronounced in the hard states, in which both systems display the so-called band limited noise and broad bumps (Figure \ref{barret_fig5}). This similarity points to a similar underlying mechanism for the variability and a similar accretion geometry in both systems. Multi-component decomposition of power density spectra using Lorentzians has shown that many of the characteristic noise frequencies (up to five is some power spectra) correlate to each other \cite{BelloniPsaltisVdKlis02}, and that in a given source, soft states are associated with higher characteristic frequencies than hard states.

As shown above, the properties of the energy spectrum can be interpreted in the framework of a truncated disk geometry with a moving inner radius. Assuming that the characteristic noise frequencies are somehow set by the inner disk radius (as expected in some theoretical models, see \cite{vanderklis00review,psaltis01}), one expects some immediate relationships between timing and spectral parameters. Consistent with this specific assumption, many clear correlations between timing and spectral parameters have been reported. One of the most significant links the amplitude of the reflection component and the low frequency quasi-periodic oscillation (QPO) detected in the power density spectra of black hole systems (see Figure \ref{barret_fig4} right). The higher the QPO frequency the stronger the reflection, as expected if indeed the QPO frequency tracks somehow the position of the inner disk radius \cite{gilfanovref-99,gilfanovref-00}. 

If considering the kilo-Hertz QPO frequencies detected in many neutron star systems, several correlations between spectral parameters and QPO frequencies have been reported (see \cite{barret01review} for a review). One correlation linked the slope of the Comptonized spectrum with the kilo-Hz QPO frequency: the steeper the slope; the higher the QPO frequency \cite{kaaret-98}. This may again be explained by the strong radiative coupling between the disk and the Comptonizing cloud invoked above to explain the correlation between the power law slope and reflection amplitude shown in Figure \ref{barret_fig4} (left).

It was also shown that kHz QPO frequencies correlate well with the luminosity on timescales of hours, but that sources differing by about three orders of magnitude in luminosity show kilo-Hz QPOs that cover the same range of frequencies \cite{vanderklis00review}. A similar effect is seen between observations of the same source at different epochs (e.g. \cite{mendez-99}). This suggests that the inner disk radius is not set by the instantaneous accretion inferred from the source luminosity. Although the physics of the truncation radius is very uncertain (see however \cite{czernyreview01}), it has been proposed that the inner disk radius could be related to some long-time averaging process over timescales of days to months \cite{vanderklisparallel-01}. In this scenario, the important parameter to set the radius would become the ratio between the total accretion rate and the average accretion rate through the disk (e.g. \cite{BO1705-03}). This scenario relies however on the uncertain assumption that the accretion rate in the disk is not the same as the total mass accretion rate. 

%\begin{figure}[!t]
%\includegraphics[width=5.50in]{como03_barret_fig6.ps}
%\caption{The picture of the accretion flow around a black hole in both hard and soft spectral states. Replacing the central black hole by a neutron star should not change too much the picture, except that the neutron star surface/boundary layer will provide an additional soft emission component to cool off the Comptonizing electrons.}
%\label{barret_fig6}
%\end{figure}

\section{Conclusions}
The spectral state transitions of X-ray binaries are more likely related to changes in the geometry of the accretion flow, which in the case of black holes is also accompanied by a change in the nature (thermal or non-thermal) of the emission. In the high state of black holes, the disk extends very close to the last stable orbit. It provides the seed photons for the Comptonization. Magnetic reconnection at the surface of the disk is the main source of heat for the electrons; a significant fraction of which have a non-thermal energy distribution. In the low state, the geometry is very much different, the disk is truncated away from the black hole, and is replaced inside by a hot optically thin inner flow. The disk which still provides the seed photons for the Comptonization does not contribute much to the X-ray spectrum, but is visible through a weak reflection component. The main weakness of that scenario is that the physics of the disk truncation is poorly constrained at the moment. This clearly means that alternative models for hard spectral states (magnetic flares or jet synchrotron emission) can still be considered. 

Neutron stars differ from black holes, not so much in terms of accretion geometry and emission process, but rather in terms of spectral behavior which is more complex, due mostly to the presence of a solid surface for the neutron star.  The unavoidable radiation from the neutron star surface acts as an additional source of cooling for the Comptonizing region, producing in general softer spectra than in the case of black holes.

Turning now to the future, thanks to its improved sensitivity in hard X-rays (at 100 keV), compared to BeppoSAX and RXTE, INTEGRAL \cite{winkler03} will enable the properties of the thermal and non-thermal Comptonized emissions from both black holes and neutron stars to be investigated in great details. These observations should however be combined with high resolution spectroscopic simultaneous observations in soft X-rays with XMM-Newton or Chandra. Following upon what has been learned recently with BeppoSAX and RXTE, one might expect significant progresses in the understanding the accretion flows around compact objects to be made in the next few years.
 
\begin{theacknowledgments} I wish to thank L. Amati, T. di Salvo, M. Gierli{\' n}ski,
M. Gilfanov, F. Frontera, E. Palazzi, R. Wijnands, A. Zdziarski for sending me their original data to make some of the figures presented in this paper. I am grateful to F. Frontera, M. Gierli{\' n}ski, M. Gilfanov, T. di Salvo for helpful comments on an earlier version of the paper. It is a real pleasure to thank the organizers of the COMO 2003 meeting, in particular D. Farina and G. Bertin for their invitation, their hospitality and their very kind support along the preparation of this paper, which arrived far too long after the original deadline.

\end{theacknowledgments}
\bibliographystyle{aipproc}   % if natbib is available

\end{document}